\documentclass[review]{elsarticle}

\usepackage{lineno,hyperref}
\usepackage{amsmath,amsthm,amssymb,mathtools,graphicx,calc,color,dsfont,psfrag}
\usepackage{array}
\usepackage{multirow}
\usepackage{pgf,tikz}
\usepackage{amsfonts}
\usepackage{bm}
\usepackage{subfigure}
\usepackage{epstopdf}
\usepackage{makeidx}
\usepackage{moreverb}
\usepackage[ruled, vlined]{algorithm2e}

\journal{Journal of Optics Communications Templates}

\begin{document}

\begin{frontmatter}

\title{Theoretical Accuracy Analysis of RSS-Based Range Estimation for Visible Light Communication}

\author[mymainaddress]{Changeez Amini}\ead{c.amini@modares.ac.ir}
\author[mymainaddress]{Paeiz Azmi\corref{mycorrespondingauthor}}
\cortext[mycorrespondingauthor]{Corresponding author}\ead{pazmi@modares.ac.ir}
\author[mysecondaryaddress]{Seyed Sadra Kashef}\ead{s.kashef@urmia.ac.ir}

\address[mymainaddress]{Department of Electrical and Computer Engineering, Tarbiat Modares University, Tehran, Iran.}
\address[mysecondaryaddress]{Department of Electrical and Computer Engineering, Urmia University, Urmia, Iran.}

\begin{abstract}
In this paper, an improved channel model of visible light communication (VLC) for ranging in presented. For indoor channel model of VLC, distance is estimated based on received signal strength. In this model, received shot noise as a  distance dependent parameter is considered in range estimation accuracy. Moreover, based on this model, the Cramer-Rao lower bound is computed as the theoretical limits on the performance and accuracy of any unbiased estimator.  In this way, the effects of horizontal and vertical distances are investigated. In addition, the transmitted power effect on RSN and accordingly on CRLB is demonstrated.
\end{abstract}

\begin{keyword}
Visible light communication,  range estimation, received shot noise, Cramer-Rao lower bound.
\end{keyword}

\end{frontmatter}



\section{Introduction}\label{sec:introduction}
Recently, visible light communication (VLC) has gradually become a research hotspot in indoor environments because of its advantages such as illumination and a high capacity for data transmission~\cite{Depth_Survey}. Additionally, the optical communication systems are safe where radio frequency (RF) is hazardous or even forbidden (same as underground mines, airplanes, submarines, intrinsically safe environments like petrochemical industries and so on). Also, the radio wave is more vulnerable to multipath effects than visible light. Furthermore, the light-emitting diode (LED) based localization is a promising approach as it can provide highly accurate positioning, inexpensively. Within the last years, increasing research in visible light positioning (VLP) can be observed which can make VLC as a more practical technology. The VLP systems use the photodiode (PD), extracting features from the received signals such as received signal strength (RSS), angle of arrival (AOA), time of arrival (TOA), and time difference of arrival (TDOA)~\cite{TDOA-based, Machine_learning, ceiling_lamp, OFDM-based}. By using the least-squares method, the accuracy of indoor visible light communication localization system based on RSS in non-line-of-sight environments is analyzed in~\cite{non-line-of-sight}. In~\cite{orientation}, the authors have investigated the simultaneous position and orientation estimation problem using RSS. Cramer-Rao lower bound (CRLB) as a fundamental limit on RSS based range estimation is investigated in~\cite{ZZB}. In~\cite{lower_bound}, the CRLB of the position error is determined using RSS measurements. It is demonstrated in~\cite{Magnitude} that RSS-based distance estimation in VLP is biased. In~\cite{low-complexityi}, a low-complexity TDOA-based VLP system using a novel and practical localization scheme based on cross-correlation has been proposed and experimentally demonstrated. Performance improvements over AOA based positioning are illustrated by simulation in which combines AOA and RSS information to enhance positioning accuracy in an asynchronous VLP system~\cite{3-D_localization}. Moreover,~\cite{16} applies the phase difference of arrival (PDOA) as the optical wireless positioning technique.

Generally, four noise sources are considered in VLP,  which are thermal noise, shot noise of background radiation, shot noise of dark current and shot noise of received signal (RSN). The total effect of these four noises is integrated as a total noise. It is assumed that the total noise has Gaussian probability distribution function (p.d.f.) and the variance of this noise equals to the sum of the variance of four noise sources.~\cite{Performance_optical}. Note that the first three noise sources are constant respect to the received signal power transmitted by LED and received in the photodetector. However, opposite to other noise sources, RSN depends on received signal power and range of receiver from the transmitter. Dependence of RSN variance to the distance affects the total noise variance considerably which can increase $\sqrt{\text{CRLB}}$ in RSS. Whereas it is shown in other fields such as channel capacity~\cite{CAPACITY, input-dependent} that the effect of input-dependent Gaussian noise (RSN) can not be ignored, this effect has not been considered so far in VLP to the best of our knowledge.

In this letter, CRLB is calculated theoretically by taking into account received signal power-dependent RSN. In this way, the effects of horizontal and vertical distances are investigated. In addition, the transmitted power effect on RSN and accordingly on CRLB is demonstrated. The rest of the paper is organized as follows: in Section \ref{sec: System Model}, we describe the system model and present the basic assumptions. In Section \ref{sec:CRLB}, we derive the exact closed-form expression for the CRLB. Simulations and related discussions are given in Section \ref{sec:simulation}. Finally, conclusions are drawn in Section \ref{sec:conclusion}.

\section{System Model and Basic Assumptions}\label{sec: System Model}
\begin{figure}[!t]
\includegraphics[height=2 in, width= 3.5 in]{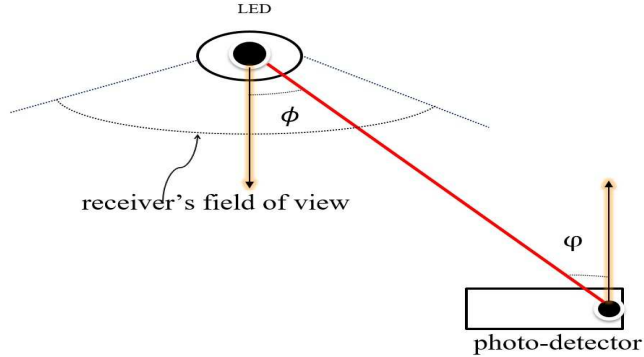}
\centering
\caption{ Two-dimensional schematic figure of the geometry of the LED source and photo-detector (receiver) in the VLC system. LOS path is the main path from the optical power source to the receiver.\label{Schematic1}}
\end{figure}
We suppose a VLC system with the corresponding  transmitter and receiver. Let $\phi$ be the angle of irradiance respect to the normal axis of transmitter surface and $\varphi$ be the angle of incidence respect to the normal axis of desk plane. Angles $\phi$ and $\varphi$ are associated with the locations of the LED source and the receiver. In this case, the channel LOS gain is given by
\begin{align}\small{
\eta_{\text{LOS}} = \left\{
\begin{array}{rl}
\frac{(m+1)S}{2\pi d^2}\,\text{cos}^m\phi \,T_{\text{s}}(\varphi)\,g(\varphi)\,\text{cos}\varphi,\quad0\le\varphi\le\varphi_\text{con}\\
0,\quad\quad\quad\quad\quad\quad\quad\quad\quad\quad\quad \varphi>\varphi_\text{con}
\end{array}, \right.}
\end{align}
where $m$ is the order of Lambertian emission, $S$ is the active area of the detector, and $d$ denotes the distance between transmitter and the receiver surface. $\varphi_\text{con}$ is the field of view (FOV) of the receiver. $T_{\text{s}}(\varphi)$ and $g(\varphi)$ are the gain of the optical filter and the gain of the optical concentrator, respectively. For sake of simplicity, the parameters are assumed such that $T_{\text{s}}(\varphi)=g(\varphi)=1$~\cite{n3}. 

In general, the multipath transmission in VLC is not challenging like RF. Typically, the power of the reflected components will be much lower than LOS path~\cite{Adaptive_threshold}. The channel gain is almost flat and the received power is given by the channel DC gain on the directed path and diffuse links~\cite{modulation_schemes}. The channel gain is given by
\begin{align}
H(0)=\eta_{\text{LOS}}+\eta_{\text{diff}},
\end{align}
where $\eta_{\text{diff}}$ expresses the channel diffuse gain. Thus the received optical power can be summarized by
\begin{align}\label{Eq-3}
x=R_\text{p}\, H(0)\, P_\text{t}+w
=P_0+P_{\text{diff}}+w,
\end{align}
where $P_0=R_\text{p}\, \eta_{\text{LOS}} \, P_\text{t}$ and $P_{\text{diff}}=R_\text{p}\, \eta_{\text{diff}} \, P_\text{t}$. $\omega$ is additive white Gaussian noise (AWGN) and $R_\text{p}$ denotes the detector responsivity. $P_\text{t}$ denotes the transmitted optical power. In practice, eye safety limitations impose a fundamental constraint on the $P_\text{t}$. 

Figure~\ref{Schematic1} illustrates the geometry of an indoor VLC system with one LED lamp located on the ceiling. The receiver has a sensor which estimates the distance between itself and the transmitter based on the measured RSS. Generally, in the VLC systems, the receiver is pointed straight up at the ceiling and the transmitter is pointed straight down from the ceiling and it is placed within the receiver-field of view. As a result, $\text{cos}\phi = \text{cos}\varphi = \frac{h}{d}$ and the LOS power can be expressed as
\begin{align}\label{Eq-4}
P_0=R_\text{p} P_\text{t}\frac{(m+1)S}{2\pi }\frac{h^{m+1}}{d^{m+3}},
\end{align}
where vertical distance is represented by $h$. Based on diffuse links considered in~\cite{x3}, the influence to CRLB of multipath reflections is ignorable when diffuse channel gain has already been measured in a specific scenario and the modulation speed is much slower than the channel cut-off frequency. Thus, $\eta_{\text{diff}}$ can be treated as a constant.

In the visible light system, the noise model could be separated into several sources as follows~\cite{Fundamental_analysis}:

\emph{1) Thermal Noise $\sigma_{\text{TH}}$:} The thermal (Johnson) noise is given as
$\sigma_{\text{T H}}=\sqrt{\frac{8\pi\kappa T_{\text{e}}}{G}\eta SI_\text{2}B^2+\frac{16\pi^2\kappa T_{\text{e}}\Gamma}{g_\text{m}}\eta^2S^2I_\text{3}B^3}$,
where $k$ denotes the Boltzmann constant, $T_{\text{e}}$ represents the absolute temperature, $G$ is the open-loop voltage gain, $\eta$ is the fixed capacitance of photodetector per unit area, $\Gamma$ is the channel noise factor, $B$ is the equivalent noise bandwidth, $g_\text{m}$ is the transconductance, and the noise bandwidth factors are $I_\text{2}= 0.562$ and $I_\text{3} = 0.0868$.

\emph{2) Shot Noise From Background Radiation $\sigma_{\text{BG}}$:} The current noise from background radiation is
$\sigma_{\text{BG}}=\sqrt{2qR_\text{p}p_{\text{BS}}S\breve{\lambda}B}$,
where $q$ is the electron charge, $p_{\text{BS}}$ is the background spectral irradiance, $\breve{\lambda}$ is the bandwidth of the optical filter.

\emph{3) Shot Noise From Dark Current $\sigma_{\text{DC}}$:} The noise from the dark current of a photodiode is
$\sigma_{\text{DC}}=\sqrt{2qI_{\text{DC}}B}$,
where $I_{\text{DC}}$ refers to the photodiode dark current.

\emph{4) RSN $\sigma_{\text{SS}}$:} The shot noise can be expressed as $\sigma_{\text{SS}}=\sqrt{2q(P_0+P_{\text{diff}})B}$. Clearly, RSN depends on LOS power. Nevertheless, RSN is supposed to be constant in previous VLP works to the best of our knowledge.

Thus the variance of Gaussian noise detected by the receiver can be summarized by
\begin{align}\label{Eq-6}
\sigma_0^2=\sigma_{\text{T H}}^2+\sigma_{\text{BG}}^2+\sigma_{\text{DC}}^2+\sigma_{\text{SS}}^2
=\sigma_{\text{T}}^2+\sigma_{\text{SS}}^2.
\end{align}
One can find a detailed discussion about noise sources and an in-depth analysis of their influence on the signal-to-noise ratio (SNR) in~\cite{Ghassemlooy}. 

\section{Cramer-Rao Lower Bound}\label{sec:CRLB}
In this part, theoretical limits on ranging accuracy are investigated by deriving the CRLB for the provided model. Note that RSS can estimate only one parameter, so it is assumed that vertical distance (height) is known which is a common assumption~\cite{x3}. First, we write the probability density function (PDF) of the received optical power using Equation (\ref{Eq-3}) as
\begin{align}\label{Eq-14}
p(x; d)=\frac{1}{\sigma_0\sqrt{2\pi}}\,\exp\left\{-\frac{1}{2\sigma_0^2}(x-P_0-P_{\text{diff}})^2\right\},
\end{align}
By using the natural logarithm, we have 
\begin{align}\label{Eq-15}
\ln p(x; d)=-\ln\sigma_0\sqrt{2\pi}-\frac{1}{2\sigma_0^2}(x-P_0-P_{\text{diff}})^2.
\end{align}
Differentiating the log-likelihood function, we have
\begin{align}\label{Eq-16}
\frac{\partial\ln p(x; d)}{\partial d}=-\frac{\frac{\partial\sigma_0}{\partial d}}{\sigma_0}+{\frac{\frac{\partial\sigma_0}{\partial d}}{\sigma_0^3}}(x-P_0-P_{\text{diff}})^2+\frac{1}{\sigma_0^2}(x-P_0-P_{\text{diff}})\frac{\partial P_0}{\partial d},
\end{align}
concerning the relationship between the RSN and the received optical power, $\frac{\partial\sigma_0}{\partial d}$ is equal to 
\begin{align}\label{Eq-17}
\frac{\partial\sigma_0}{\partial d}=qB\big(\sigma_{\text{T}}^2+2q(P_0+P_{\text{diff}})B\big)^{-\frac{1}{2}}\frac{\partial P_0}{\partial d}
=\frac{qB}{\sigma_0}\frac{\partial P_0}{\partial d}.
\end{align}
Alternatively, by using Equation (\ref{Eq-17}), we can rewrite Equation (\ref{Eq-16}) as
\begin{align}\label{Eq-18}
\frac{\partial\ln p(x; d)}{\partial d}=-\frac{qB}{\sigma_0^2}\frac{\partial P_0}{\partial d}+\frac{qB}{\sigma_0^4}(x-P_0-P_{\text{diff}})^2\frac{\partial P_0}{\partial d}+\frac{1}{\sigma_0^2}(x-P_0-P_{\text{diff}})\frac{\partial P_0}{\partial d}.
\end{align}
By using Equation (\ref{Eq-17}), the second derivative would be
\begin{align}\label{Eq-11}
&\frac{\partial^2\ln p(x; d)}{{\partial d}^2}=\frac{2(qB)^2}{\sigma_0^4}(\frac{\partial P_0}{\partial d})^2
-\frac{qB}{\sigma_0^2}\frac{\partial^2P_0}{{\partial d}^2}-4\frac{(qB)^2}{\sigma_0^6}(x-P_0-P_{\text{diff}})^2(\frac{\partial P_0}{\partial d})^2\nonumber\\
&-\frac{2qB}{\sigma_0^4}(x-P_0-P_{\text{diff}})(\frac{\partial P_0}{\partial d})^2+\frac{qB}{\sigma_0^4}(x-P_0-P_{\text{diff}})^2\frac{\partial^2P_0}{{\partial d}^2}\nonumber\\
&-2\frac{qB}{\sigma_0^4}(x-P_0-P_{\text{diff}})(\frac{\partial P_0}{\partial d})^2-\frac{1}{\sigma_0^2}(\frac{\partial P_0}{\partial d})^2+\frac{1}{\sigma_0^2}(x-P_0-P_{\text{diff}})\frac{\partial^2P_0}{{\partial d}^2}.
\end{align}
Upon taking the negative expected value we have
\begin{align}\label{Eq-14}
-\mathbb{E}\bigg[\frac{\partial^2\ln p(x; d)}{{\partial d}^2}\bigg]=(\frac{\partial P_0}{\partial d})^2[\frac{2(qB)^2}{\sigma_0^4}+\frac{1}{\sigma_0^2}].
\end{align}
Then, the variance of any unbiased estimator must satisfy
\begin{align}\label{Eq-38}
\sqrt{\text{var}(\widehat{d_0})}\ge\sqrt{\frac{1}{\sigma_0^2+2(qB)^2}}\,\frac{\sigma_0^2}{|\frac{\partial P_0}{\partial d}|},
\end{align}
where $|\frac{\partial P_0}{\partial d}|$ refers to absolute value of $\frac{\partial P_0}{\partial d}$ which is given by
\begin{align}\label{Eq-39}
\frac{\partial P_0}{\partial d}=-\frac{R_\text{p}P_\text{t}S}{2\pi }(m+1)(m+3)\frac{h^{m+1}}{d^{m+4}}.
\end{align}
\section{Numerical Results and Discussion}\label{sec:simulation}
In this section, the simulation results are provided to evaluate the performance of the RSS method. The CRLB results (by taking account of the impact of the channel DC gain, as well as the RSN,) are simulated and studied in this section. The simulations parameters are listed in Table~\ref{tabel1}.
\\
\begin{table}[t]
\centering
\caption{Some typical parameter values}
\begin{tabular}[b]{l l @{} c}
\hline
Parameter&
\multicolumn{2}{c}{$\quad\quad$ Value} \\
\hline
\\
$\text{Electron Charge}~(q)
$ &&\quad\quad$1.6\times10^{-19}$~\text{C} \\
$\text{Dark Current}~(I_\text{DC})
$ &&\quad\quad$5$~\text{pA} \\
$\text{Channel Noise Factor}~(\Gamma)
$ &&\quad\quad$1.5$\\
$\text{Photodiode Responsively}~(R_\text{p})
$ &&\quad\quad$0.4$~\text{mA/mW} \\
$\text{Equivalent Noise Bandwidth}~(B)
$ &&\quad\quad$400$~\text{MHz}\\
$\text{Open-Loop Voltage Gain}~(G)
$ &&\quad\quad$10$\\
$\text{Boltzmann's Constant}~(\kappa)
$ &&\quad\quad$1.38\times10^{-23}$\\
$\text{Absolute Temperature}~(T_\text{e})
$&&\quad\quad$300$~\text{K}\\
$\text{Active Area}~(S)
$&&\quad\quad$.2~\text{cm}^2$\\
$\text{Transconductance}~(g_\text{m})
$ &&\quad\quad$30$~\text{ms}\\
$\text{Optical Filter Bandwidth}~(\breve{\lambda})
$&&\quad\quad$400\;(380-780)$~\text{nm} \\
$\text{Spectral Irradiance }~(p_\text{BS})
$ &&\quad\quad$5.8\times10^{-6}$~\text{W/cm$^2\cdot$nm} \\
$\text{Photodiode Fixed Capacitance}~(\eta)
$ &&\quad\quad$112$~\text{pF/cm$^2$} \\
$\text{Interference}~(P_{\text{diff}})
$&&\quad\quad0~\text{W} \\
\\
\hline
\end{tabular}
\label{tabel1}
\end{table}

\begin{figure}[!t]
\includegraphics[height=2 in, width= 3.5 in]{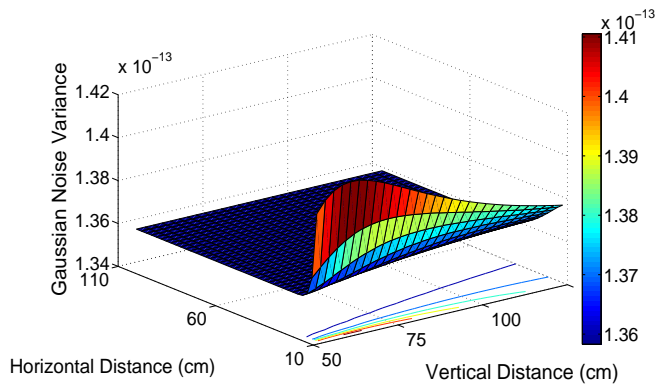}
\centering
\caption{The influence on the variance of Gaussian noise of RSN for a typical room environment.\label{variance2}}
\end{figure}

By considering the RSN, the Gaussian noise variance consists of a constant parameter ($\sigma_{\text{T}}^2$) and a variable one ($\sigma_{\text{SS}}^2$). According to the parameters of Table~\ref{tabel1}, we can calculate $\sigma_{\text{TH}}^2=1.12\,10^{-13}$, $\sigma_{\text{BG}}^2=2.38\,10^{-14}$, and $\sigma_{\text{DC}}^2=6.40\,10^{-22}$ which in total is $\sigma_{\text{T}}^2=1.36\,10^{-13}$. The value of $m$ can be expressed as $m = -\frac{\text{ln}2}{\text{ln}(\theta_\frac{1}{2})}$, where $\theta_\frac{1}{2}$ is defined as the view angle when irradiance is half of the value at $0^\circ$. When the value of $m$ increases, it means the value of $\theta_\frac{1}{2}$ decreases. The signal is more concentrated and the SNR is higher. Therefore, the accuracy is higher. By considering this fact, CRLB is plotted in~\cite{x3} versus source optical power with several numbers of $m$. The effect of the variable parameter is shown in Fig.~\ref{variance2} which is a contour plot under the three-dimensional shaded surface where $m=50$, and $P_t=1 \text{W}$. Obviously, the effect of the RSN is mainly visible just below the LED, while further away (horizontally) from the LED, the effect is diminishable. Although, because of benefits provided by higher SNRs, it is more likely to utilize the receiver near to the transmitter. If so, RSN ($\sigma_{\text{SS}}^2$) is not negligible and has a noticeable effect on the analysis of indoor VLC system.

\begin{figure}[!t]
\includegraphics[height=2 in, width= 3.5 in]{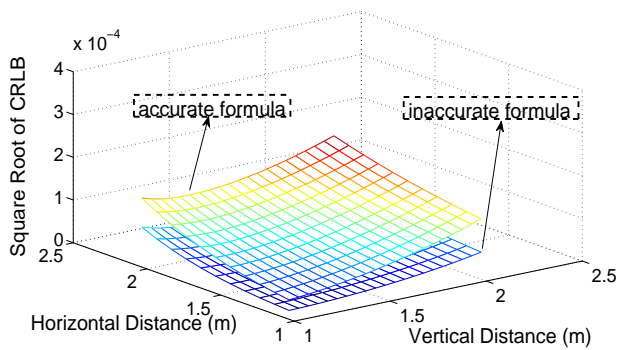}
\centering
\caption{$\sqrt{\text{CRLB}}$ for proposed accurate formula by  Equation (\ref{Eq-38}) against inaccurate one (equation (16) of~\cite{x3}) as a common formula in literature. \label{crlb6}}
\end{figure}

In Fig.~\ref{crlb6}, the effect of distance over $\sqrt{\text{CRLB}}$ is shown where it is assumed $m=1$, $P_t=1 \text{W}$. $\ell$ refers to the horizontal distance between the transmitter and receiver and $\ell=\sqrt{d^2-h^2}$. $\ell$ varies from 1 to 2 meters, and $h$ varies from 1 to 3 meters. If the receiver is placed at a larger distance, $\sqrt{\text{CRLB}}$ will increase. More importantly, this figure represents that under the new model, the $\sqrt{\text{CRLB}}$ is higher than what was previously supposed. This considerable difference originates from the nature of VLC noise model, illustrating that RSN can't be ignored and has a significant effect on the theoretical accuracy analysis of VLP. As mentioned before, the other noise components are assumed to be more dominant in the literature. We emphasize that without solving the CRLB in detail, the relation between RSN and CRLB could not be found.

\begin{figure}[!t]
\includegraphics[height=2 in, width= 3.5 in]{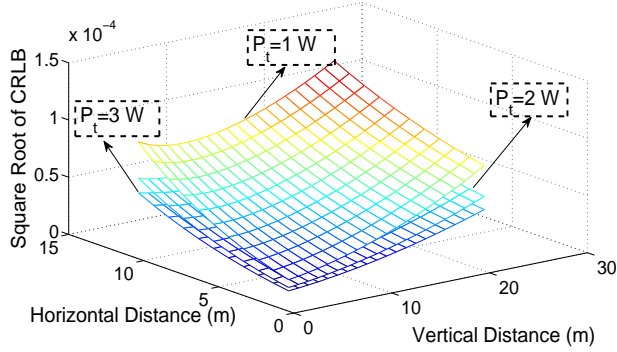}
\centering
\caption{$\sqrt{\text{CRLB}}$ for proposed accurate formula by  Equation (\ref{Eq-38}) with several transmitted optical powers. \label{photonics1}}
\end{figure}

In Fig.~\ref{photonics1}, for several values of $P_t$, CRLB is plotted where it is assumed $m=1$, $\ell$ varies from 1 to 2 meters, and $h$ varies from 1 to 3 meters. Obviously, by increasing the transmitted optical power, the CRLB decreases toward lower values. More significantly, this figure shows that the relation between the transmitted optical power and the CRLB is so complicated. $\sqrt{\text{var}(\widehat{d_0})}$ is proportional to $\frac{1}{P_t}$ for lower values of received optical power, whereas for higher values of received optical power, $\sqrt{\text{var}(\widehat{d_0})}$ is proportional to $\frac{1}{\sqrt{P_t}}$. In other words, CRLB will decrease for higher values of received optical power. Although increasing the $P_t$ is one way to rise the received optical power, eye safety regulations will limit the amount of transmitted optical power. However, received optical power could be risen by some other ways such as increasing $R_p$ or $S$, reducing $d$, and utilizing tilting technique. 

\begin{figure}[!t]
\includegraphics[height=2 in, width= 3.5 in]{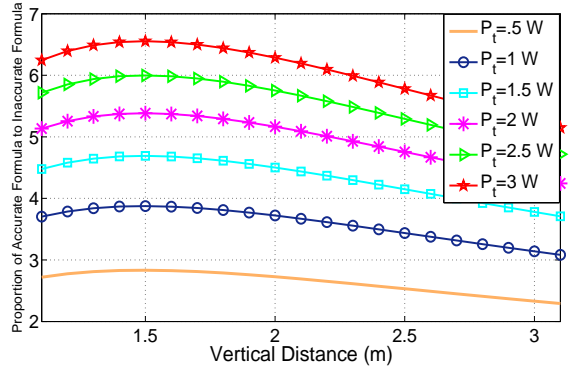}
\centering
\caption{The proportion of accurate formula to inaccurate formula with several transmitted optical powers. \label{photonics2}}
\end{figure}

The proportion of accurate formula of $\sqrt{\text{CRLB}}$ to inaccurate one is shown in Fig.~\ref{photonics2}. The figures present the mean of $\sqrt{\text{CRLB}}$ with regard to horizontal distance where $m=1$, $\ell$ varies from 1 to 2 meters, and $h$ varies from 1 to 3 meters. It can be clearly seen from Fig.~\ref{photonics2} that the figures grow by increasing $P_t$. However, the rate of growth declines for higher values of transmitted optical power. This is because the impact of $P_t$ on CRLB will decrease sharply for higher values of received optical power. In addition, Fig.~\ref{photonics2} illustrates that the relation between the transmitted optical power and the CRLB is complicated which should be discussed in the future.

\begin{figure}[!t]
\includegraphics[height=2 in, width= 3.5 in]{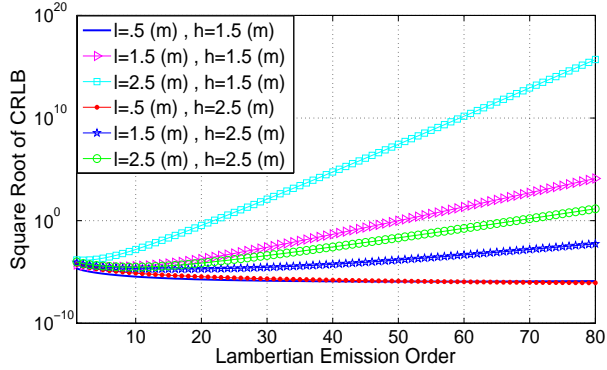}
\centering
\caption{$\sqrt{\text{CRLB}}$ against the order of Lambertian emission with several values of horizontal distance and vertical distance. \label{crlbcombined_new}}
\end{figure}

In Fig.~\ref{crlbcombined_new}, the value of $\sqrt{\text{CRLB}}$ is evaluated by varying the horizontal distance and vertical distance. The radiation model of LEDs is not perfect Lambertian distribution in practical applications and $m$ might be different in different LED manufacturers. We vary the value of $m$ from $1$ to $80$. Fig.~\ref{crlbcombined_new} indicates that by increasing horizontal distance for the given height, $\sqrt{\text{CRLB}}$ rises where $P_t=1 \text{W}$. Moreover, this figure shows that $\sqrt{\text{CRLB}}$ is a convex function, i.e. $\sqrt{\text{CRLB}}$ has a minimum value where $m=m_\text{opt}$. It is proved in~\cite{time-of-arrival} that $m_\text{opt}\approx -(2+\frac{1}{\text{lncos}\phi})+\sqrt{1+(\frac{1}{\text{lncos}\phi})^2}$. In the future works, we would calculate $m_\text{opt}$ more accurately.

\section{Conclusion}\label{sec:conclusion}
For an indoor positioning system based on visible light, we have presented an analysis of the CRLB of RSS-based ranging. In this work, RSN is considered in modelling and showed that can not be negligible. By taking into account RSN and its relation to distance parameter, a more accurate noise model and closed-form expression for CRLB are achieved. A detailed discussion about the parameters that determine the CRLB and their dependence on the geometry of the system is also presented. In addition, we have investigated the effect of Lambertian emission order on the $\sqrt{\text{CRLB}}$ and showed that $\sqrt{\text{CRLB}}$ is a convex function.

\end{document}